\documentclass[review]{elsarticle}

\usepackage{lineno,hyperref}
\usepackage{amsmath}
\modulolinenumbers[5]
\journal{Journal of \LaTeX\ Templates}

\begin{document}

\begin{frontmatter}

\title{A wide swing charge sensitive amplifier for a prototype Si-W EM calorimeter}


\author[mymainaddress,mysecondaryaddress]{Sourav Mukhopadhyay\corref{mycorrespondingauthor}}
\cortext[mycorrespondingauthor]{Corresponding author}
\ead[url]{souravm@barc.gov.in}

\author[mymainaddress,mysecondaryaddress]{Vinay B. Chandratre}
\author[mymainaddressa,mysecondaryaddress]{Sanjib Muhuri}
\author[mymainaddressa]{Rama N. Singaraju}
\author[mymainaddressa]{Jogender Saini}
\author[mymainaddressa,mysecondaryaddressb,mysecondaryaddressc]{Tapan K Nayak}

\address[mymainaddress]{Bhabha Atomic Research Centre, Trombay, Mumbai -- 400085, India}
\address[mymainaddressa]{Variable Energy Cyclotron Centre, Kolkata -- 700097, India}
\address[mysecondaryaddress]{Homi Bhabha National Institute, HBNI, Mumbai -- 400094, India}
\address[mysecondaryaddressb]{National Institute of Science Education and Research, Jatni -- 752050, India}
\address[mysecondaryaddressc]{CERN, Geneva 23, Switzerland}

\begin{abstract}
A wide swing charge sensitive amplifier (CSA) has been developed, as a part of a front-end electronics (FEE) readout ASIC, for a prototype silicon tungsten (Si-W) based electromagnetic (EM) calorimeter. The CSA, designed in 0.35 $\mu$m N-well CMOS technology using 5V MOS transistors, has a wide linear operating range of 2.6 pC w.r.t the input charge with a power dissipation of 2.3 mW. A noise figure (ENC) of 820 e- at 0 pF of detector capacitance with a noise slope of 25 e-/pF has been achieved (when followed by a CR-RC$^2$ filter of 1.2 $\mu$s peaking time). This design of CSA provides a dynamic range (ratio of maximum detectable signal to noise floor) of 79 dB for the maximum input charge of 2.6 pC when connected to a silicon detector with a capacitance of 40 pF. Using folded cascode architecture-based input stage and low voltage high swing current mirrors as the load, the CSA provides an enlarged output swing when biasing the output node towards one supply rail and utilizing the voltage range towards the opposite rail.  The design philosophy works for both polarities of a large input signal. This paper presents the design of CSA with a wide negative output swing for an anticipated input signal of positive polarity in the target application with a known detector biasing scheme. 
\end{abstract}

\begin{keyword}
\texttt CSA \sep wide swing FEE \sep  charge preamplifier
\end{keyword}

\end{frontmatter}


\section{Introduction}
\label{sec:intro}

The continuous advancement in the field of particle accelerators for the next generation of high energy physics (HEP) experiments ~\cite {hep-1} puts forward significant challenges of particle identification, separation, and analysis. Advanced HEP detectors like highly granular calorimeters ~\cite {hep-2, hep-3, hep-4} demand high-density readout FEE with a wide dynamic range (about 80 dB) and a precision better than 1\% ~\cite {hep-1}. Hence, the first stage preamplifier design in the readout FEE chain is critical and should satisfy the dynamic range requirements with adequate signal to noise ratio (SNR) in a power-efficient way to enable high-density integration.

As the first stage of FEE, the CSA is preferred primarily because of its low power, low noise configuration and charge gain's (A$_Q$) insensitivity to a variation of detector capacitance (C$_D$). For Si-W based EM calorimeter, which involves particle shower formation and dissolution, a wide dynamic range (simultaneous detection of maximum energy deposition $\approx$ few pC in the shower maximum layers and detection of MIP particle, i.e. $\approx$ 4 fC  for a silicon detector of 300$\mu$m wafer thickness~\cite {mip}) CSA is required as the first stage of FEE. 

Various design approaches to address these critical requirements are reported in the literature. A switchable gain charge preamplifier is one of them with multiple feedback capacitances to improve the output swing ~\cite {switch-gain}. Here, the gain variation has to be done either through slow peripheral control ~\cite {slow-periphery} or through additional gain selection circuitry involving comparator~\cite {casis}. The latter option may introduce non-linear behaviour in the circuit during switching among the various gain modes. Two separate charge preamplifier channels with charge division/sharing methods using series capacitance is another reported design scheme ~\cite {taku}. However, this method requires one extra channel per detector element, leading to extra power dissipation, additional area consumption, a compromised low signal performance and increasing cost. A combination of switchable gain CSA with ADC+ToT (Time over Threshold) technique is also reported ~\cite {hgroc} to extend the swing of the FEE. But, this method has faced few challenges like high cross-talk, long dead time, absence of smooth overlap between the ADC and ToT methods when the preamplifier goes from the non-saturation mode to saturation mode ~\cite {hgroc-ppt}.

A wide swing CSA can facilitate all these design methods to achieve an even higher linear operating range. This paper thus reports a design technique to enhance the linear operating range of the charge preamplifier itself by optimizing the output dc point close to one supply rail while utilizing the voltage range toward the opposite supply rail as output swing. Section II presents the architecture of the CSA. The analysis for the major design parameters is provided in section III. Section IV describes the design methodology for the CSA. Simulation results and the layout view of the CSA are reported in section V. Section VI presents the test results of the fabricated and packaged CSA as a part of a bigger FEE ASIC. Finally, section VII concludes the paper.

\section{Architecture: The CSA}
\label{sec:archi}

\begin{figure}[htbp]
\centering 
\includegraphics[width=1.0\textwidth]{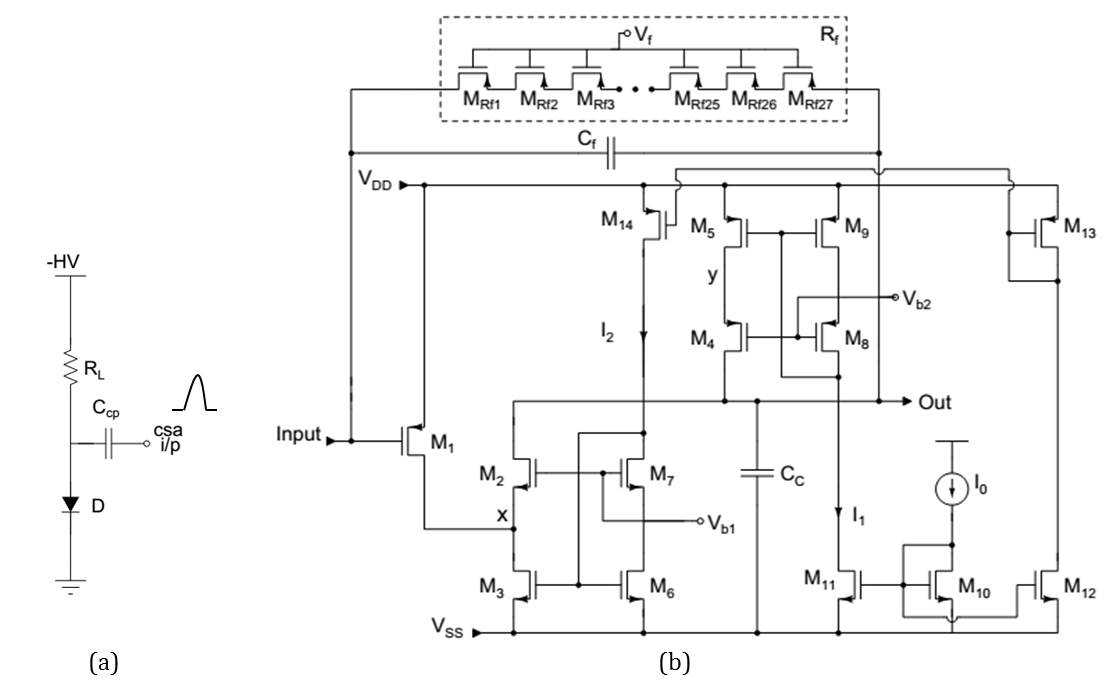}
\caption{\label{csa-bias} (a) Detector biasing scheme in the target application, (b) Schematic of the CSA.}
\end{figure}

The known detector biasing scheme in the target application for this CSA is shown in Figure~\ref {csa-bias}a, where R$_L$ is the detector bias resistance, D represents a P-N junction (diode) detector, and C$_{cp}$ is the ac coupling capacitor between the detector and the CSA. In this biasing scheme with an N-type silicon PAD detector, the anode (p-type side) and cathode (n-type side) of the detector are connected to the negative HV and ground. As a result, the expected signal polarity at the output of the CSA is negative, and as it involves one polarity inversion, the equivalent signal polarity at the input will be positive, as shown in Figure~\ref {csa-bias}a. 

The CSA has been designed in 0.35 $\mu$m CMOS technology using 5V MOS transistors with a typical V$_{thn}$=0.75 V, V$_{thp}$=-1.0 V, and k$_n$=100 $\mu$A/V$^2$, k$_p$=30 $\mu$A/V$^2$. In this design, the CSA is optimized to have the output dc bias point close to the positive supply rail ($\approx$ 4 V) and utilize the voltage swing towards the negative supply rail to allow a large negative signal swing. The threshold voltage of the input transistor (M$_1$) puts the limit on the max output dc voltage. A schematic of the proposed CSA is shown in Figure~\ref {csa-bias}b. The architecture of the CSA is based on three sections; a core amplifier stage (M$_{1-9}$), a feedback network (C$_f$ (an integrating capacitor) and R$_f$ (a decay resistor)), and the biasing network (M$_{10-14}$). 

The core amplifier is designed using a folded cascode based input stage and low voltage high swing current mirror as the load. Folded cascode architecture facilitates high gain and wide bandwidth design with a minimum number of active components with enhanced output swing and ease of frequency compensation. Low voltage high swing current mirror architecture is chosen as the active load to satisfy the broader dynamic range requirement of the CSA. It reduces the voltage headroom requirement from 2V$_{dsat}$+V$_{th}$ of the cascode current mirror to 2V$_{dsat}$ apart from providing very high output resistance (of the order of g$_m$r$_{ds}^2$) ~\cite {razavi}. Assuming a conservative value of average V$_{dsat}$ also, this architecture improves the output voltage swing to be wide enough into the supply rails ($\sim$ V$_{DD}$ – 4V$_{dsat}$). 

The value of C$_f$ is a trade-off between the charge/conversion gain (A$_Q$) and rise time (shown in ~\ref{sec:A}) of the CSA. This work aims to design a wide dynamic range CSA, so C$_f$ was optimized to be 0.8 pF  providing a A$_Q$  of 1.25 mV/fC (lower A$_Q$ leads to enhanced linear range w.r.t. input charge). The dc operating point is set by shunt-shunt feedback using a very high-value R$_f$ (to reduce the thermal noise contribution 4kT/R$_f$). R$_f$ is implemented by a series of transistors (M$_{Rf1-27}$) operating in the linear/triode region with its gate voltage (V$_f$) externally controlled to have a variable R$_f$ (2 - 120 MΩ) and adjustable fall time. Series connection of multiple transistors ensures low distortion and improved linearity over a large signal swing. In this design, V$_f$ is externally adjusted for an equivalent feedback resistance of 75 MΩ providing a fall time constant of 60 $\mu$s for the CSA output response, which again sets a limit in rate capability to $\approx$ 10 kHz.

The biasing network, consisting of transistor M$_{10-14}$, distributes the required current (I$_1$ and I$_2$) from a master current source (I$_0$) through current mirroring action. V$_{b1}$ and V$_{b2}$ are optimized bias voltages for maximum swing. C$_c$ is the compensating capacitor at the high impedance output node of the CSA to improve the stability response.

\section{Design analysis}
\label{sec:analysis}

In this section, the analysis of the major design parameters (noise and ac performance) of the CSA is performed.

\subsection{Noise optimization}
The primary noise contributors in a typical detector and front-end readout configuration (as shown in Figure ~\ref{csa-detector} in ~\ref{sec:A}) are the shot noise due to the detector leakage current and the electronic noise of the CSA. The expression for effective electronic noise at the input of the core amplifier (considering only the thermal noise contribution of the transistors) of Figure~\ref {csa-bias}b is

\begin{equation}
\label{eq1}
S_{vin,th}(f) = \frac{\frac{8kT}{3} \cdot (g_{m1} + g_{m3} + g_{m5})}{g_{m1}^2}
\end{equation}

Where g$_m$’s are the trans-conductance of the respective transistors, k is the Boltzmann constant, and T is the temperature in Kelvin. By ensuring higher g$_{m1}$, the effective input noise will be dominated by M$_1$. 
Then, the total input-referred noise spectral density (including the 1/F noise) of the core amplifier can be expressed as ~\cite {sansen},

\begin{equation}
\label{eq2}
S_{vin,tot}(f)=\frac{8kT}{3g_{m1}} + \frac{K_{1/f}}{c_{ox}^2 WLf}
\end{equation}

Where g$_{m1}$ is the trans-conductance of the M$_1$, C$_{ox}$ is the oxide capacitance per unit area, K$_{1/f}$ is the flicker noise coefficient, and WL is the area of the input transistor. Equation ~\ref{eq2} shows that the type of the input transistor, its working region, and operating parameters play a vital role in determining the noise level. In this work, the M$_1$ is chosen to be PMOS as it is advantageous over NMOS regarding 1/F noise in submicron technology ~\cite {oconner}. A very high aspect ratio (10,0000)  is used (discussed in section~\ref{sec:method}) for the M$_1$ with a low bias current (I$_d$) of 350 $\mu$A to maximize the transistor g$_m$ efficiency (g$_m$/I$_d$ $\approx$ 20), with low current density (I$_d$/W $\approx$ 0.07 A/m) signifying the operation of the transistor in the sub-threshold region at the edge of weak/moderate inversion ~\cite {subth1, subth2}. 

\subsection{ac analysis}

\begin{figure}[htb]
\centering 
\includegraphics[width=0.9\textwidth]{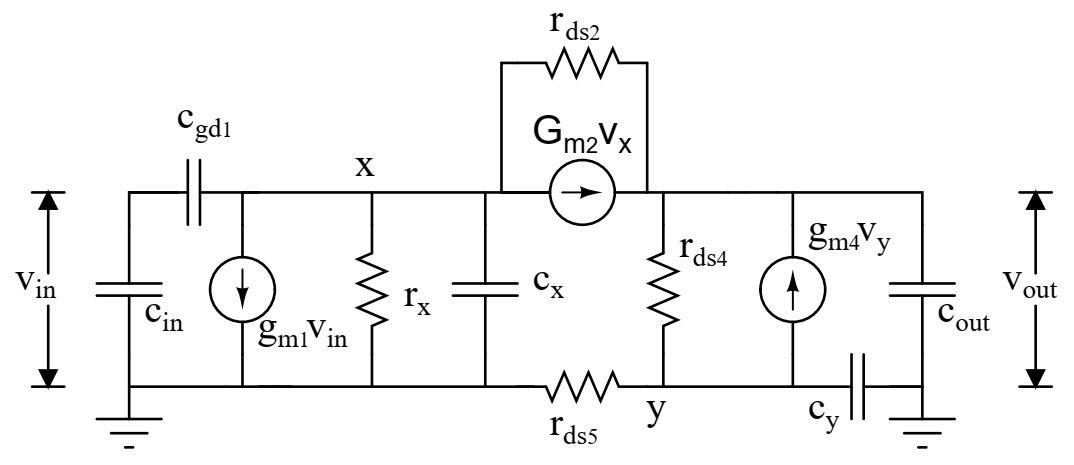}
\caption{\label{small-signal} Simplified small-signal equivalent model of the core amplifier.}
\end{figure}

The simplified small-signal equivalent model of the core amplifier is shown in Figure~\ref {small-signal}, where, c$_{in}$ = c$_{gs1}$, r$_x$= r$_{ds1}||r_{ds3}$, c$_x$= c$_{gd3}+c_{gs2}+c_{gd1}+c_{db1}$, G$_{m2}$= g$_{m2}+g_{mb2}$, c$_y$= c$_{gd5}+c_{gs4}$, c$_{out}$ = c$_c+c_{gd2}+c_{gd4}$. Now ignoring the very high-frequency pole at node y, the above small-signal model can be approximated with a second-order system where the dominant pole is at the output high impedance node while the non-dominant pole is accompanied with the folded node (x). 

Hence the following expressions can be derived. The open-loop dc gain,

\begin{equation}
\label{eq3}
A_{v0} = g_{m1} \cdot r_{out} 
\end{equation}

The low frequency pole (p$_1^*$) at the output,
\begin{equation}
\label{eq4}
 p_1^* = - \frac{1}{2\pi \cdot r_{out} \cdot c_{out}}                                                                        
\end{equation}

Where, r$_{out}$ = $(g_{m2} \cdot r_{ds2} \cdot (r_{ds1}|| r_{ds3})) || (g_{m4} \cdot r_{ds4} \cdot r_{ds5})$

The high frequency (p$_2^*$) second pole,

\begin{equation}
\label{eq5}
 p_2^* = - \frac{1}{2\pi} \cdot \frac {G_{m2}}{c_x}                                                                                                                                                                                           
\end{equation}

The resultant UGB (Unity Gain Bandwidth) of the core amplifier,

\begin{equation}
\label{eq6}
 UGB(f_0) = A_{v0} \cdot p_1^* = - \frac{1}{2\pi}  \cdot \frac {g_{m1}}{c_{out}}                                                         
\end{equation}

\section{Design methodology}
\label{sec:method}
The low power budget of 2 mW, defined by the application, roughly sets the maximum bias current for the CSA to be 400 $\mu$A. The authors have set I$_1$ to be 50 $\mu$A, and I$_2$ to be 400 $\mu$A and reasonable V$_{dsat}$ for each transistor according to the bias current, voltage swing, and transistor specific operating region requirements. Since M$_3$ carries the maximum current, it has been assigned a larger V$_{dsat}$ of 500 mV, while the cascode transistor M$_2$ has been provided with 200 mV of V$_{dsat}$. V$_{dsat}$ of 300 mV has been set for M$_4$ and M$_5$ in the current mirror load due to the lower mobility of the PMOS transistors. V$_{b1}$ and V$_{b2}$ are biased at V$_{th,M2} + V_{dsat3}+ V_{dsat2}$ and V$_{DD}$ - V$_{th,M4} - V_{dsat4} - V_{dsat5}$ respectively to ensure the output swing of the core amplifier is V$_{DD}$ - V$_{dsat5} - V_{dsat4} - V_{dsat3} - V_{dsat2}$. A very low V$_{dsat}$ of 50 mV has been considered for the input transistor M$_1$, to obtain a high g$_m$ value with a low bias current.  Aspect ratios are calculated based on the equation shown in Table~\ref {tab:1}, which are later optimized to achieve the high dc loop gain (A$_{v0}$\textgreater80 dB) for maximum charge transfer from the detector to the CSA, low input impedance (Z$_{in}$), and subsequently low crosstalks among the adjacent channels~\cite {sampa-thesis}. Minimum allowed transistor length (L$_{min}$) is used for the input transistor (M$_1$) to ensure maximum g$_m/C_{gs}$ (”Figure of Merit”) ~\cite {oconner, tech-doc}. The cascode transistor (M$_2$) is designed with a higher aspect ratio to place the non-dominant pole of the core amplifier away from the 3-dB frequency of the closed-loop system of CSA coupled with a detector, discussed in section ~\ref{sec:calc} and ~\ref{sec:A}. M$_3$ and M$_5$ are designed with larger L and higher V$_{dsat}$ to reduce their noise contribution. The calculated and the final aspect ratio of the transistors are given in Table~\ref {tab:1}. An aspect ratio of 0.5 is kept for the feedback transistors (M$_{Rf1}-M_{Rf27}$) to ensure they work in the linear region. 

\begin{table}[htb]
\centering
\caption{\label{tab:1}Aspect ratio of the transistors in the core amplifier}
\begin{tabular}{c c c c c c c}
\hline \hline  
Transistor & \vtop{\hbox{\strut I$_d$}\hbox{\strut ($\mu$A)}}  & \vtop{\hbox{\strut {K$_{eff}$=$\mu c_{ox}$}}\hbox{\strut ($\mu$A/V$^2$)}} & \vtop{\hbox{\strut V$_{dsat}$}\hbox{\strut (V)}}  & \vtop{\hbox{\strut Calculated (W/L)}\hbox{\strut =2$\cdot$I$_d$/(K$_{eff}\cdot$V$_{dsat}^2$)}} & \vtop{\hbox{\strut Final}\hbox{\strut (W/L)}} & \vtop{\hbox{\strut Operating}\hbox{\strut Region$^*$}} \\ \hline
M$_1$ & 350 & 30 & 50m & 10000 & 10000 & 3 \\
M$_2$ & 50 & 100 & 200m & 25 & 100 & 2 \\
M$_3$ & 400 & 100 & 500m & 32 & 50 & 2 \\
M$_4$ & 50 & 30 & 300m & 40 & 50 & 2 \\
M$_5$ & 50 & 30 & 300m & 40 & 50 & 2 \\ \hline
\end{tabular}
\footnotesize {$*$3: Sub-threshold region (edge of weak/moderate inversion), 2: Saturation region.}
\end{table}

Table~\ref {tab:2} provides the dc bias voltages at different nodes in the circuit.
 
\begin{table}[htbp]
\centering
\caption{\label{tab:2}Bias voltages at different nodes in the circuit}
\begin{tabular}{c c c}
\hline \hline
Sr Number & Node & dc bias point (V) \\ \hline
1. & Input & 3.985 \\
2. & x & 0.42 \\
3. & Out & 3.985 \\
4. & y & 4.5 \\
5. & V$_{b1},V_{b2}$ & 1.5, 3.2 \\ \hline
\end{tabular}
\end{table}

\subsection{Design calculation}
\label{sec:calc}
In this design for the core amplifier, with the value of g$_{m1}\approx$ 7 mS, c$_{out}\approx$ 4.5 pF, the value of f$_0$ is $\approx$ 247 MHz. According to ~\ref{eq4} and ~\ref{eq5}, the low-frequency pole (p$_1^*$) of the core amplifier is at 1.8 kHz, while the high-frequency pole (p$_2^*$) is at 50 MHz with r$_{out}\approx$ 20 MΩ, c$_x\approx$ 3.5 pF, and G$_{m2}\approx$ 1.1 mS. The calculated open-loop dc gain is 103 dB.
The second-order transfer function of the closed-loop system incorporating the CSA coupled with a silicon detector (as shown in ~\ref{sec:A}), involves two poles  p$_{rfcf}$ due to the feedback network (R$_f$ and C$_f$) and p$_{cfb}$ due to the network involving C$_f$, C$_t$, the core amplifier. The location of these two poles can be calculated as (considering negligible load at the output), 

\begin{equation}
\label{eq7}
p_{rfcf}=\frac {1}{2\pi \cdot R_f \cdot C_f} = 2.7 \:\text {kHz}                                                                                                                                                                                
\end{equation}

\begin{equation}
\label{eq8}
p_{cfb}=\frac{f_0 \cdot C_f}{C_t} = 4.4 \:\text {MHz}    
\end{equation}

Where, with detector capacitance of 40 pF, C$_t$ $\approx$ 45 pF. The closed-loop bandwidth (-3dB frequency) is evaluated by Equation ~\ref{eq8}, which is also the unity gain frequency of the charge feedback loop. The dominant pole (Equation ~\ref{eq7}) is around 2.7 kHz, and the non-dominant pole (p$_2^*$) sits at a frequency around 50 MHz. The overall phase margin of the system is then given by (Equation ~\ref{eq9}).

\begin{equation}
\label{eq9}
PM = tan^{-1}(\frac {p_2^*}{p_{cfb}}) = 85 \:\text {degree}                                                                                      
\end{equation}



\section{Simulation results}
\label{sec:sim}

\begin{figure}[htbp]
\centering 
\includegraphics[width=1.0\textwidth]{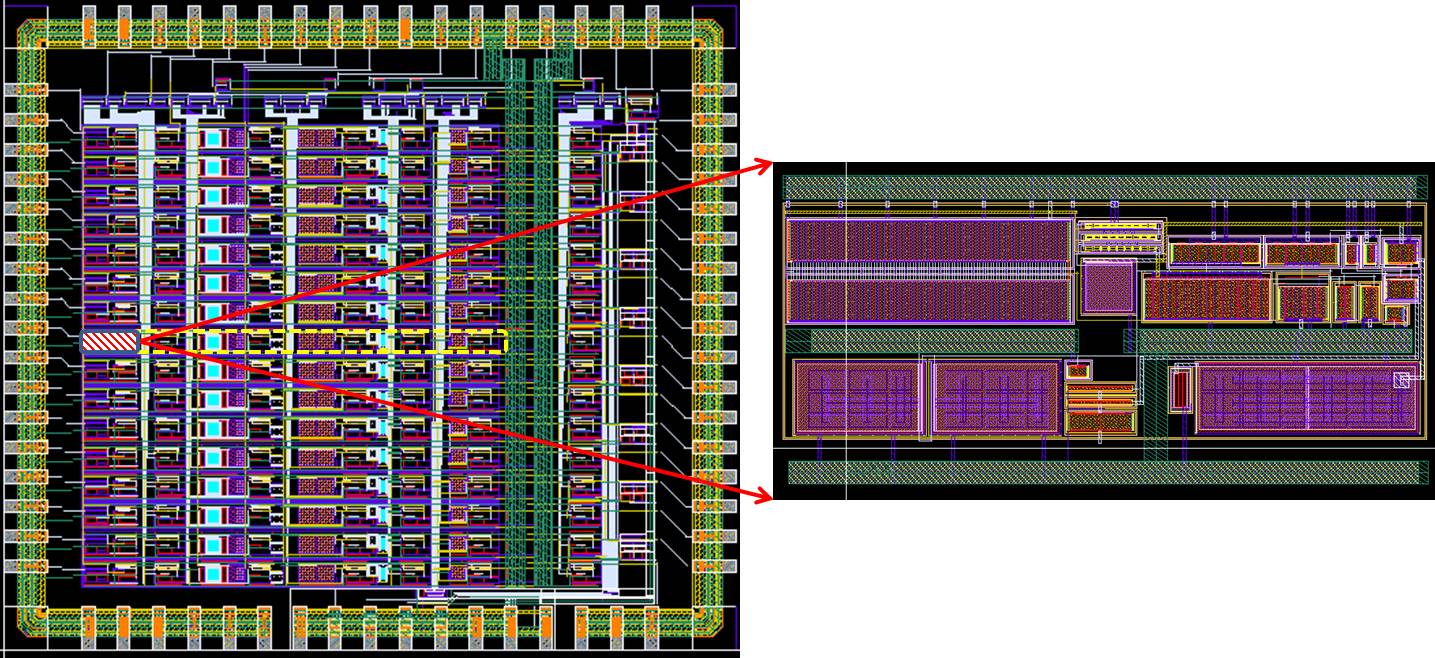}
\caption{\label{layout} Layout view of the 16 channel pulse processing ASIC highlighting a single channel (yellow dotted) and CSA block (red hashed) along with a zoomed layout view of the CSA.}
\end{figure}

The physical design of the CSA has been carried out using an inter-digitized layout for the large input transistor~\cite {razavi}, using multiple contacts, frequent substrate contact, and guard rings to minimize the parasitic noise due to the resistive poly gate and distributed substrate resistance. Figure~\ref {layout} shows the layout view of the designed CSA (415 $\mu$m x 200 $\mu$m) as a part of the 16 channel pulse processing ASIC (5.6 mm x 5.2 mm) developed as FEE for Si-W calorimeter. 

\begin{figure}[hbt!]
\centering 
\includegraphics[width=0.9\textwidth]{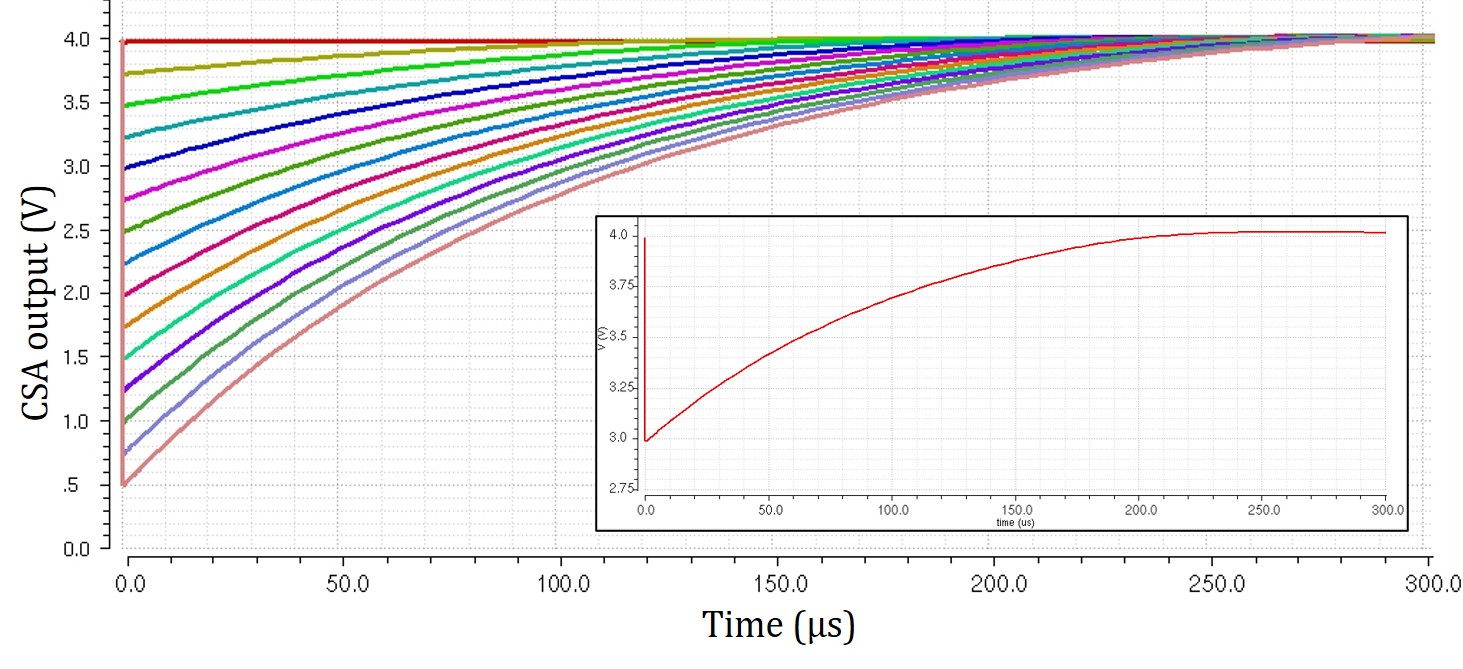}
\caption{\label{transient} Transient response of the CSA with linear behaviour up to a test charge of 2.8 pC with a zoomed view in the inset showing response with 0.8 pC of input charge.}
\end{figure}

The simulations of the CSA were carried out with the silicon extracted view across all the design corners provided by the foundry. To inject a test charge in the simulation, a tail pulse (using an exponential voltage pulse) was used with a test (coupling) capacitor (0.8 pF to have the closed-loop voltage gain of 1) connected in series with the CSA input. The transient response at the CSA output is shown in Figure~\ref {transient}. The output dc point of the CSA is biased at 3.985 V and a linear behaviour of the CSA circuit can be seen up to a injected test charge of $\approx$ 2.8 pC (corresponds to input voltage of 3.5 V into a 0.8 pF coupling capacitor). 

Table~\ref {tab:3} summarizes the salient design features of the simulated CSA with a 40 pF detector capacitance (C$_{det}$).

\begin{table}[hbt!]
\centering
\caption{\label{tab:3} Simulated characteristics of the CSA}
\begin{tabular}{c c c}
\hline \hline
Sr. No & Parameter & Value \\ \hline  
1. & dc Loop gain & 83.73 dB \\
2. & Phase margin & 86 degree \\
3. & Unity gain frequency & 3.33 MHz \\
4. & Power consumption & 2.3 mW \\
5. & ENC$^*$ (C$_{det}$ = 0 pF) & 500 e$^-$ \\
6. & Noise slope$^*$ & 18 e$^-$/pF \\
7. & Conversion gain & 1.25 mV/fC \\
8. & Area & 415 $\mu$m x 200 $\mu$m \\
9. & Technology & 0.35 $\mu$m CMOS \\
10. & Supply & +5V, GND \\  \hline
\end{tabular}
\\
\footnotesize {$*$ when followed by a CR-RC$^2$ filter of 1.2 $\mu$s peaking time}
\end{table}

\section{Test Results}
\label{sec:test}

\begin{figure}[hbt!]
\centering 
\includegraphics[width=1\textwidth]{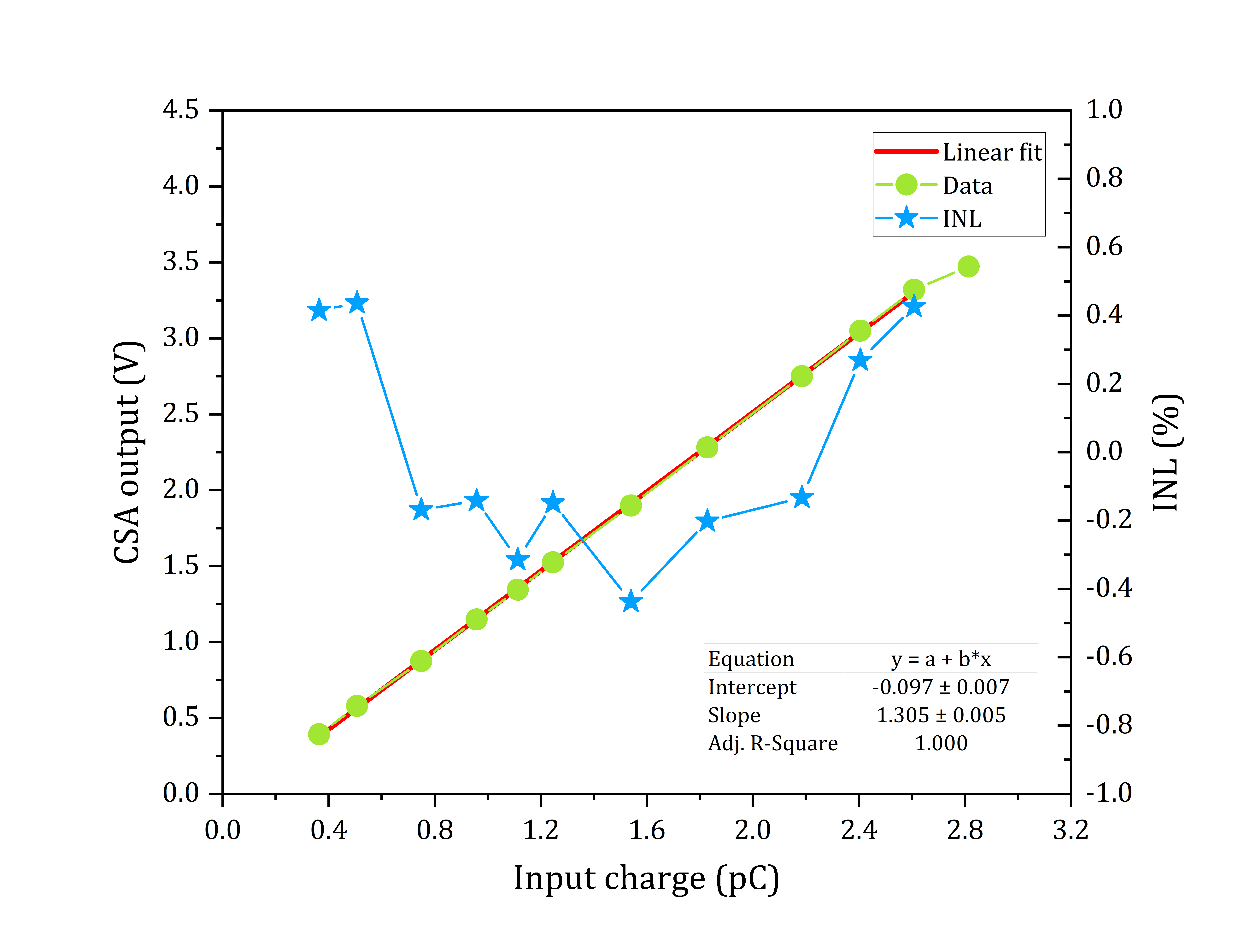}
\caption{\label{linearity-residue} The linear response of the CSA.}
\end{figure}

As a part of a bigger FEE ASIC, the CSA was fabricated in a 0.35 $\mu$m CMOS technology and packaged in CLCC 68. The FEE ASIC consists of 16 pulse processing channels, where each channel consists of CSA, semi-gaussian pulse shaper, track \& hold and gain stage. For diagnostic and calibration purposes, all the individual stage output of a particular channel was taken out to the package pins. The fabricated CSA has been functionally tested utilizing the calibration channel in the packaged ASIC in the laboratory. The linearity of the wide swing CSA has been measured by observing the output response on the oscilloscope with increasing input voltage pulse in series with a coupling capacitor (to inject charge) and plotted in Figure~\ref {linearity-residue}. The linear response of the CSA is satisfactory with an Integral Non Linearity better (INL) than $\pm$0.4\% up to 2.6 pC of input charge (corresponding to an output swing of 3.3 V). The measured conversion/charge gain is $\approx$ 1.3 mV/fC, while the ENC and noise slope shows a higher value of 820 e$^-$ and 25e$^-$/pF respectively from that achieved in the simulation. This results in a dynamic range of 79 dB. The cross-talk measured at the final output of the FEE ASIC is $\sim$ 0.6\%. Table~\ref {tab:4} summarizes the test results of the CSA.

\begin{table}[hbt!]
\centering
\caption{\label{tab:4} Performance summary of the CSA}
\begin{tabular}{c c c}
\hline \hline
Sr. No & Parameter & Value \\ \hline  
1. & ENC$^*$ (C$_{det}$ = 0 pF) & 820 e$^-$ \\
2. & Noise slope$^*$ & 25 e$^-$/pF \\
3. & INL & better than $\pm$0.4\% up to 2.6 pC\\
4. & Dynamic range & 79 dB \\
5. & Conversion gain & 1.3 mV/fC \\
6. & Cross-talk$^{**}$ & 0.6\% \\ \hline
\end{tabular}
\\
\footnotesize {$*$ when followed by a CR-RC$^2$ filter of 1.2 $\mu$s peaking time}\\
\footnotesize {${**}$ Measured at the final output of the FEE ASIC}
\end{table}

\begin{figure}[hbt!]
\centering 
\includegraphics[width=0.6\textwidth]{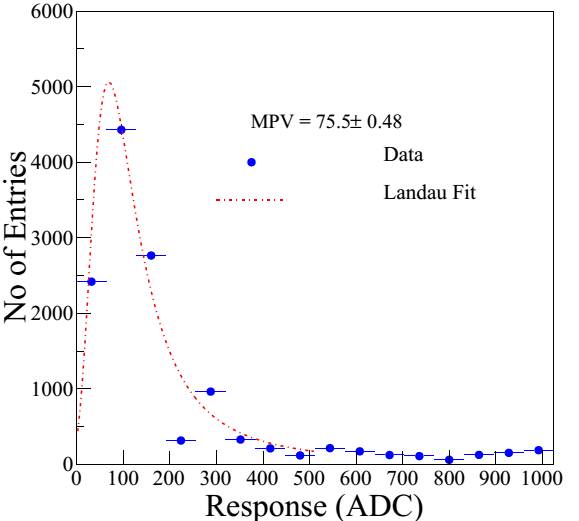}
\caption{\label{mip} Minimum ionizing particle (MIP) response of a silicon pixel of the prototype FOCAL, facing the beam and readout by the FEE ASIC. The data for MIP particle is fitted with a landau distribution and resulted in 75 ADC counts as the most probable value (MPV) of energy loss (with a calibration factor of 15 used during analysis, corresponds to $\approx$ 5 mV) as expected.}
\end{figure}

The ASIC has been validated at the SPS beamline, CERN as FEE readout for the prototype forward calorimeter (FOCAL) ~\cite {focal}, a proposed electromagnetic calorimeter as part of the ALICE upgrade. The response of the prototype calorimeter was linear up to incident energy of 90 GeV for an electron beam validating the wide swing FEE. The MIP response, measured with a 120 GeV of pion beam, is shown in Figure~\ref{mip}, which signifies the low noise performance of the ASIC. These results, as depicted in Figure~\ref {linearity-residue} and Figure~\ref{mip}, validate the wide dynamic range CSA, of this paper, as the first stage of the FEE ASIC.

\section{Conclusion}
\label{sec:conclude}

A wide swing CSA has been designed in a 0.35 $\mu$m CMOS technology, fabricated, and tested. The architecture of the CSA is based on a folded cascode amplifier as the input stage and wide swing current mirrors as the load. The CSA, designed with a reasonably low power budget of 2.3 mW, has shown a linear response up to $\approx$ 2.6 pC of input charge with a dynamic range of 79 dB. The design has been fabricated as a part of a 16 channel pulse processing FEE ASIC. The ASIC was successfully validated at the SPS beamline, CERN, as a FEE readout for the prototype FOCAL detector, using a 120 GeV pion beam and up to 90 GeV of an electron beam. For an anticipated input signal of positive polarity in the target application for a known detector biasing scheme, this design has achieved an enlarged negative output swing by biasing towards one supply rail and utilizing the voltage range towards the other supply rail. Similarly, for the opposite polarity (negative) of a large input signal, the same design architecture reported in this paper can be utilized using the complementary types of transistors (shown in ~\ref{sec:B}). Moreover, the combination of the wide swing CSA of this paper and the ADC+ToT ~\cite {hgroc} technique can result in FEE with an even higher dynamic range (shown in ~\ref{sec:C}).

\section {Acknowledgements}
The authors are thankful to Bhabha Atomic Research Centre, Mumbai -- 400085, and Variable Energy Cyclotron Centre, Kolkata -- 700097 India, for supporting the research work. The authors would also like to thank all the crew members of the SPS beamline for the excellent quality of beam for the detector test and ALICE-FOCAL collaboration for the support during the tests.

\appendix
\section{The CSA and the silicon detector}
\label{sec:A}
\begin{figure}[hbt!]
\centering 
\includegraphics[width=0.6\textwidth]{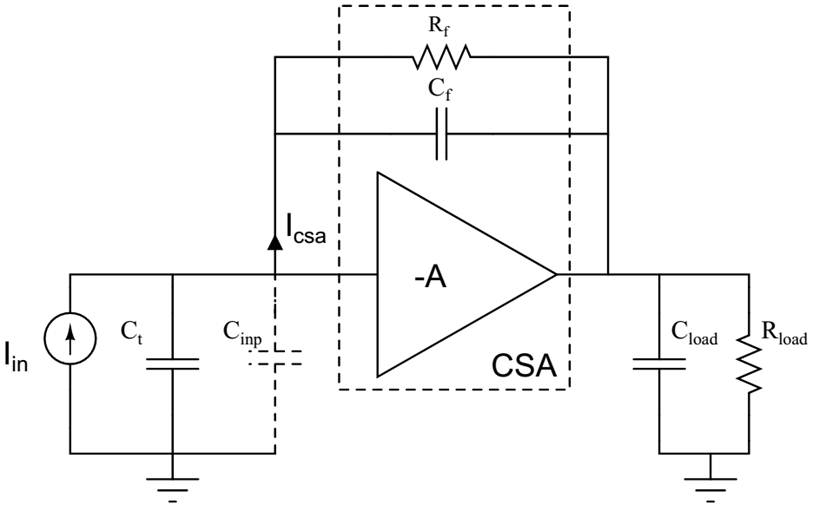}
\caption{\label{csa-detector} Typical CSA and silicon detector configuration.}
\end{figure}

The generalized circuit configuration of the CSA with silicon detector at the input and the output load is shown in Figure~\ref {csa-detector}. The total signal charge Q generated inside the detector (due to signal current I$_{in}$(t)) will be distributed between the total input capacitance C$_t$ (C$_{det}$+C$_f$+C$_{gs1}$+C$_{gd1}$) ~\cite {sansen} and the dynamic input impedance C$_{inp}$ of the CSA. The dynamic input impedance of the CSA will be resistive and C$_{inp}$ = 1/($\omega_0$$\cdot$C$_f$) ~\cite {spieler}, where $\omega_0$=2$\pi$f$_0$ and f$_0$ corresponds to the unity-gain bandwidth of the core amplifier with the output load. The expression for I$_{csa}$, the current getting integrated across the C$_f$ can be written as,

\begin{equation}
\label{eqA.1}
I_{csa}(s) = I_{in} (s) \cdot \frac{\frac{1}{sC_t}}{\frac{1}{sC_t} + \frac{1}{\omega_0 \cdot C_f}}
               = I_{in}(s) \cdot \frac{\omega_0 \cdot C_f}{sC_t + \omega_0 \cdot C_f}
               = I_{in}(s) \cdot \frac{1}{\tau_2} \cdot \frac{1}{s+1/\tau_2} 
\end{equation}

Where, $\tau_2$=  C$_t$/($\omega_0$ $\cdot$ C$_f$). Now,  

\begin{equation}
\label{eqA.2}
V_{out} (s) = I_{csa}(s) \cdot \frac {R_f}{1+sC_f \cdot R_f}
                 = I_{in}(s) \cdot \frac {1}{\tau_2} \cdot \frac{1}{s+\frac{1}{\tau_2}} \cdot \frac {1}{C_f}  \cdot \frac{1}{s+\frac{1}{\tau_1}}
\end{equation}

Where, $\tau_1$= R$_f$ $\cdot$ C$_f$. Rearranging,

\begin{equation}
\label{eqA.3}
\frac{V_{out}(s)}{I_{in}(s)}=\frac {1}{\tau_2} \cdot \frac{1}{s+\frac{1}{\tau_2}} \cdot \frac {1}{C_f}  \cdot \frac{1}{s+\frac{1}{\tau_1}}
\end{equation}

Equation ~\ref{eqA.3} gives the transfer function of the closed-loop system involving the detector and the CSA. The transfer function involves two time constants ($\tau_1$ and $\tau_2$), one due to the feedback network (R$_f$, C$_f$) and the other due to the network involving C$_f$, C$_t$, the core amplifier. 
Since the detector current can be approximated as a Dirac impulse with an integrated area of Q, the Laplace transform of I$_{in}$(t) is equal to Q. Hence, it can be written as,

\begin{equation}
\label{eqA.4}
V_{out}(s)= \frac {Q}{C_f \cdot \tau_2} \cdot \frac {1}{(s+\frac{1}{\tau_1}) \cdot (s+\frac{1}{\tau_2})}
\end{equation}

Taking the inverse Laplace transform,
\begin{equation}
\label{eqA.5}
V_{out}(t)=\frac {Q \cdot \tau_1}{C_f \cdot (\tau_1-\tau_2)} \cdot [e^{-\frac {t}{\tau_1}} - e^{-\frac{t}{\tau_2}}]
\end{equation}                                                                                                       

The expression of Equation ~\ref{eqA.5} is the general form of the time-domain response of the CSA output, where A$_Q$ = 1/C$_f$ is the charge gain of the CSA.
$\tau_1$ and $\tau_2$ are the time constants associated with the two poles p$_{rfcf}$ and p$_{cfb}$. The value of the pole p$_{cfb}$ (= ($\omega_0$$\cdot$C$_f$)/(2$\pi$$\cdot$C$_t$)) happens to be equal to the unity gain frequency of the charge feedback loop. So, to ensure stability, the high-frequency poles of the core amplifier (p$_2^*$) need to be designed away from p$_{cfb}$. Equation~\ref{eqA.5} shows the fall time is determined by the feedback resistor and capacitor (R$_f$$\cdot$C$_f$), while the rise time is inversely proportional to the product of f$_0$  and C$_f$ and directly proportional to C$_t$. 

\section{Wide swing CSA for the opposite polarity of an input signal}  
\label{sec:B} 
\begin{figure}[hbt!]
\centering 
\includegraphics[width=1.0\textwidth]{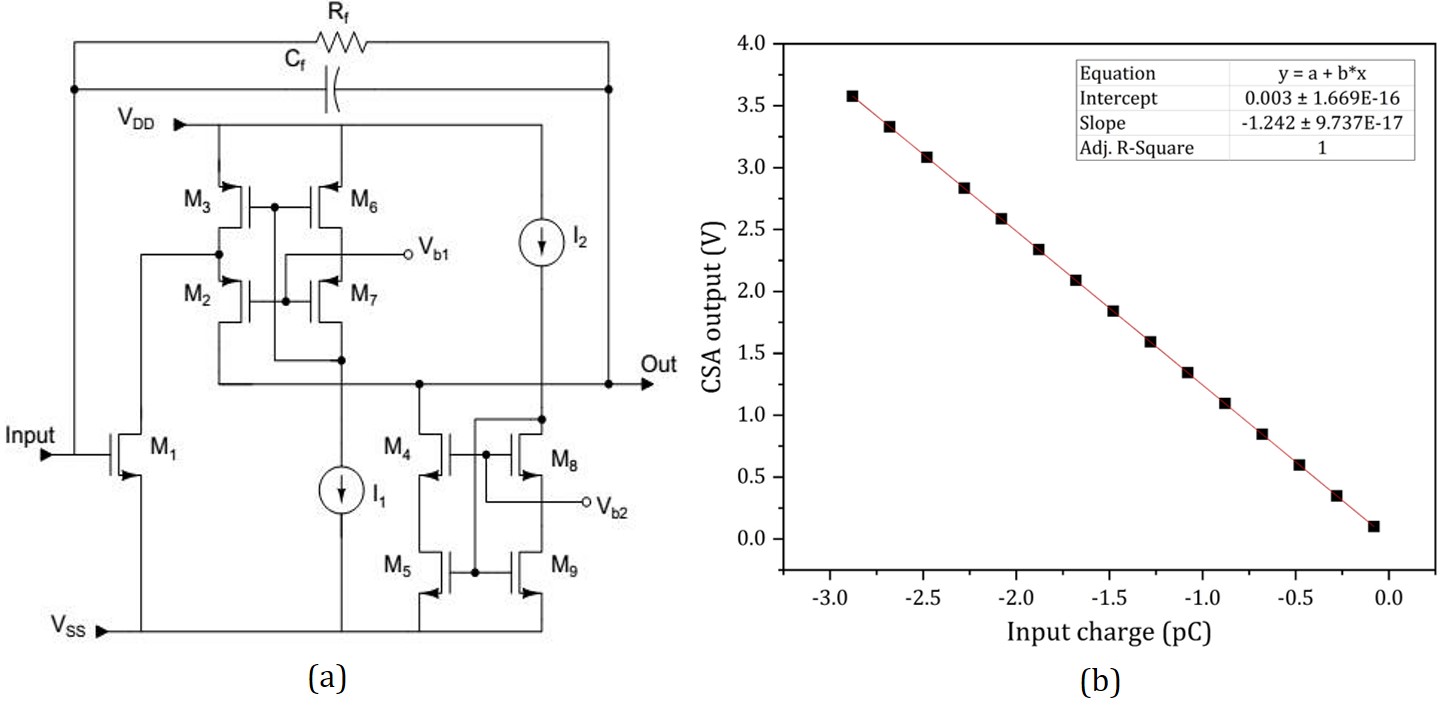}
\caption{\label{appendix-b} The simplified schematic of the wide swing CSA for a negative input signal.}
\end{figure}

Using the same architecture as shown in section 2, the design of CSA can be modified to accommodate a large negative swing of the input signal. All the transistors of Figure~\ref {csa-bias}b need to be changed to their complementary type respectively. Figure~\ref {appendix-b}a shows a simplified schematic of the same. The linear performance of the circuit has been plotted in Figure~\ref {appendix-b}b. 

\section{Wide swing CSA with ToT to further improve the dynamic range}  
\label{sec:C}                                                                                                                                                
\begin{figure}[hbt!]
\centering 
\includegraphics[width=0.7\textwidth]{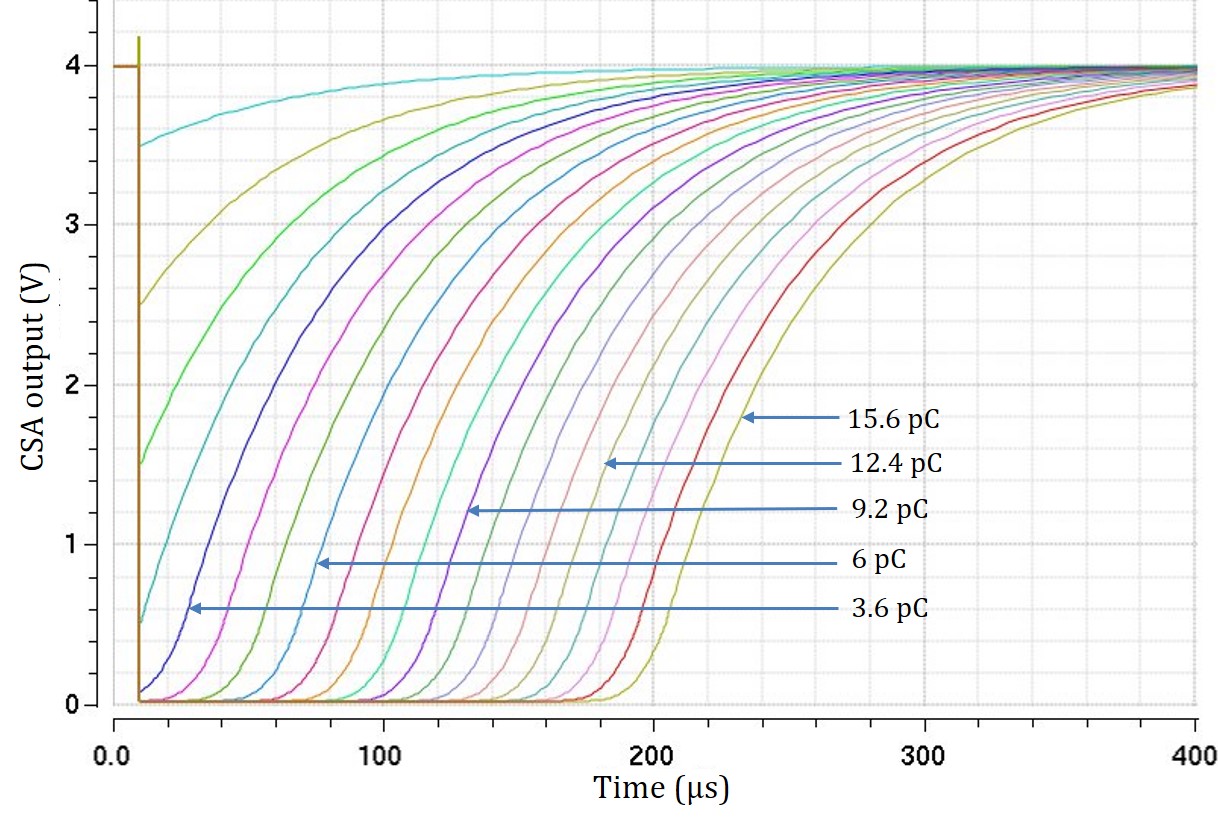}
\caption{\label{appendix-ca} Transient response of the CSA with gradually increasing input signal; The falling edge of the transient response can be seen to be proportionally spaced in time once the CSA output pulse height is saturated.}
\end{figure}

As a proof-of-concept of the dynamic range enhancement technique with the combination of the ToT scheme and the conventional pulse amplitude analysis using ADC, the designed CSA of this paper has been simulated with a much larger amplitude of the input signal than the original linear range of 2.6 pC. Figure ~\ref {appendix-ca} shows the transient response of the CSA with a parametric sweep of the input signal from low to very high amplitude. 

\begin{figure}[hbt!]
\centering 
\includegraphics[width=0.7\textwidth]{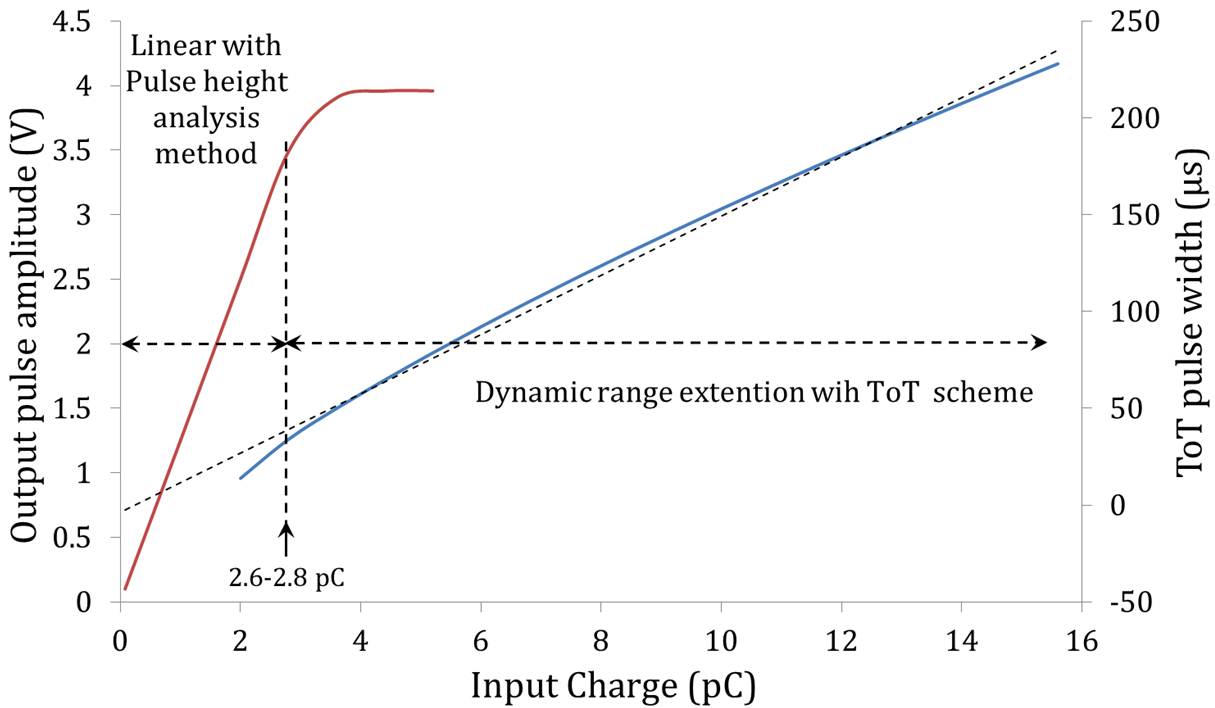}
\caption{\label{appendix-cb} Dynamic range enhancement of the CSA with a combination of pulse amplitude analysis and ToT scheme. It can be seen that up to an input signal of 2.6-2.8 pC, the pulse amplitude (red line) of the CSA output is proportional whereas, with a higher input signal, the time over threshold pulse width (blue line) is shown proportional behaviour.}
\end{figure}

Time over threshold pulse width has been measured from Figure ~\ref {appendix-ca} by keeping a threshold voltage of 2V (where the CSA output gets saturated over 3.5V) and plotted in Figure ~\ref {appendix-cb}. It can be interpreted that the linear operating range of the CSA is enhanced up to more than 15 pC with reasonable linearity.

\bibliography{csa-ref}

\end{document}